\documentclass[]{spie}  
\usepackage[]{graphicx}
\usepackage{textgreek}
\usepackage{hyperref}
\usepackage{enumitem}
\usepackage{amsmath}

\title{SimCADO - an instrument data simulator package for MICADO at the E-ELT} 

\author{
K.~Leschinski\supit{a}, 
O.~Czoske\supit{a},
R.~K\"ohler\supit{b,a},
M.~Mach\supit{a},
W.~Zeilinger\supit{a},
G.~Verdoes Kleijn\supit{c}
J.~Alves\supit{a}, 
W.~Kausch\supit{a,b},
N.~Przybilla\supit{b},\\
the A*~consortium\footnote{\: \:The A* consortium comprises the Astronomy and Astrophysics departments at: - the University of Vienna - Universit\"at Innsbruck - Johannes Kepler University Linz - the University of Graz - in collaboration with the Johann Radon Institute for Computational and Applied Mathematics (RICAM)},
and the MICADO~consortium\footnote{\: \:The MICADO consortium has partners in Germany (the Max Planck Institute for extraterrestrial Physics (PI institute), the Max Planck Institute for Astronomy, the Observatory of the University of Munich, and the Institute for Astrophysics of the Georg-August University), France (CNRS/INSU, represented by LESIA), the Netherlands (NOVA, with specific contributions to MICADO from the University of Gronigen, the University of Leiden, and the NOVA optical/infrared instrumentation group), Austria (the A* consortium), and Italy (INAF, represented by the Observatory of Padova).}
\skiplinehalf
\supit{a}
Institute for Astrophysics, 
University of Vienna, 
T\"urkenschanzstra\ss e 17, 
1180 Vienna, 
Austria; \\
\supit{b}
Institut f\"ur Astro- und Teilchenphysik, 
Universit\"at Innsbruck,
Technikerstra\ss e 25/8, 
6020 Innsbruck, 
Austria; \\
\supit{c}
Kapteyn Astronomical Institute,
University of Groningen,
Postbus 800, 
9700 AV Groningen,
The Netherlands
}

\authorinfo{For further information please contact Kieran Leschinski: \href{mailto:kieran.leschinski@univie.ac.at}{kieran.leschinski@univie.ac.at}}

\begin{document} 
  \maketitle 

\begin{abstract}
MICADO will be the first-light wide-field imager for the European Extremely Large Telescope (E-ELT) and will provide diffraction limited imaging (7mas at 1.2\textmu m) over a $\sim$53 arcsecond field of view. In order to support various consortium activities we have developed a first version of SimCADO: an instrument simulator for MICADO. SimCADO uses the results of the detailed simulation efforts conducted for each of the separate consortium-internal work packages in order to generate a model of the optical path from source to detector readout. SimCADO is thus a tool to provide scientific context to both the science and instrument development teams who are ultimately responsible for the final design and future capabilities of the MICADO instrument. Here we present an overview of the inner workings of SimCADO and outline our plan for its further development.
\end{abstract}


\section{INTRODUCTION}
\label{sec:intro}  


The global astronomical community is currently embarking on an ambitious journey to construct the largest optical and near-infrared telescopes the world has ever seen. The European Southern Observatory (ESO) has recently put into motion plans to build the largest of these next generation telescopes. The European Extremely Large Telescope (E-ELT)\cite{eelt}, with its 39m primary mirror and intrinsic adaptive optic systems, aims to observe the faintest and most distant objects in the universe at the diffraction limit. MICADO\cite{micado} - the {\bf M}ulti-AO {\bf I}maging {\bf CA}mera for {\bf D}eep {\bf O}bservations - will be the near-infrared (NIR) wide-field camera available at the E-ELT at first-light. 

MICADO will be the work-horse instrument for imaging observations. It will provide diffraction limited images in the near infrared (NIR) wavelength range over a field of view (FoV) of $\sim$53$\times$53 arcsec$^2$ in wide-field mode by using an array of nine 4096 x 4096 pixel detectors with a plate-scale of 4 mas. A zoom mode, with a plate-scale of 1.5 mas, will also be provided for a FoV of $\sim$ 16$\times$16 arcsec$^2$. Additional observation modes also include a long slit spectrograph with R$\sim$4000, windowed high time resolution imaging and high contrast imaging. Observing at the diffraction limit will be facilitated by either a Single Conjugate Adaptive Optics (SCAO) system or MAORY's Multi-Conjugate Adaptive Optics (MCAO) system.

As the complexity of observatories and instruments increases, so too has the need to predict exactly how instruments will perform. In order to achieve this goal, the current generation of instrumentation projects are investing time and effort in developing instrument data simulators during their design phases (e.g. PhoSim for the LSST\cite{phosim}, TOAD for 4MOST\cite{4most}, HSIM for HARMONI\cite{zieleniewski2015}, the METIS simulator\cite{schmalzl2012}). The MICADO consortium is no different. We are currently in the process of completing SimCADO, an instrument data simulator for MICADO. SimCADO is a python package that collects and combines the efforts of the various consortium-internal work packages and provides a framework for simulating raw output images based on the most recent design of the instrument. SimCADO does not aim to simulate each and every photon along its path from the source to the read out electronics, instead SimCADO transforms the incoming photon flux by using array and matrix operations which best represent the spectral and spatial effects generated by each optical element along the optical train.

This paper is organised in the following manner: Section 2 provides background to the SimCADO project and outlines the intended user groups. Section 3 gives an overview of how SimCADO models and applies the effects that each of the optical components has on the incoming photon flux. The data model behind SimCADO, and how the user interacts with the package is described in Section 4. This section also discuss the limitations of SimCADO as well as outline the plans for future updates to the package.

\section{SimCADO, the data simulation effort for MICADO} 
\label{sec:SimCADO}


SimCADO is a data simulation package written for Python 3 with the goal of generating realistic mock detector plane array read out files for MICADO and the E-ELT. It allows the user to see the effect that the optical train has on the incoming photons by providing the user with raw data sets that will be similar to what MICADO will produce during a typical observing run. SimCADO is also highly configurable. The user is able to simulate various observational scenarios, e.g. the use of different adaptive optics (AO) systems, or set the effectiveness of different subsystems along the optical train, e.g. the performance of the derotator or atmospheric dispersion corrector (ADC). These aspects of SimCADO are covered in greater detail in Sec. \ref{sec:modelling}. 

\subsection{SimCADO user groups}

Besides simulation related deliveries towards ESO, the idea for such an instrument data simulator was born out of the needs of various work-package teams within the MICADO consortium, namely the:

\begin{description}
    \item Science team
    
    The task of the science team is to define and develop the primary science drivers for the MICADO instrument. In order to better determine the feasibility of these science cases, the team requires a tool with which to consistently simulate calibrated images for each case. 

	\item Data reduction pipeline team
	
	In order to begin the development of the data reduction pipeline before the instrument enters the construction phase, the data reduction pipeline team (also known as the data flow system (DFS) team) will need sets of simulated raw images from the detector array. These test data sets can be provided by an instrument data simulator. With such data the team will test the performance of algorithms to correct for the instrumental fingerprint and determine physical calibration for the instrument.
	
	\item Instrument design team
	
	To create MICADO many work packages are required, which in turn means a large number of sub-systems need to be modelled in detail. However modelling the entire optical train at the level of detail used for each sub system would require a prohibitive effort. A tool which takes the results of each sub-system modelling effort and uses these to quickly model the whole optical train - from source to read out - will be useful for deciding between hardware solutions and software solutions for correcting instrumental effects.

	\item Instrument control software team
	
	Knowing quantitatively how various parameters affect the image quality is crucial for designing the software needed to drive the instrument and for determining how much observing time to allocate for various types of targets.
	
\end{description}

SimCADO aims to meet all these needs. In the immediate future, SimCADO will be the tool that the Science and Instrument design teams use as they iterate on the design of the optical train. By allowing both teams to visualise the effect of different designs on the detector array image, a design for MICADO can be chosen which best balances the desires of the science goals with the feasibility of hardware solutions.

It should also be noted that the MICADO consortium is required to provide data for an Exposure Time Calculator (ETC) to ESO, which will be made available to the astronomical community. The ETC provides the community with basic quantitative information regarding the capabilities of MICADO and is the first stop when testing out the feasibility of new ideas for observations. SimCADO will provide the quantitative information on the capabilities of MICADO needed to power the ETC.

Looking further into the future, we intend for the astronomical community to also benefit from SimCADO. As time on the E-ELT will be in high demand and difficult to obtain, it is important that the astronomical community be able to plan the best possible observing strategies. Being able to predict the outcomes of different schemes and techniques will assist in achieving the highest scientific return once the E-ELT is on line. The ability to simulate what will be possible in 10 years will also allow the astronomical community to lay down road-maps for the next decade as well as conduct the ground work needed to best understand these future observations.

\subsection{Design requirements for SimCADO}

Balancing the needs of the various user groups is not an easy task. For example, while each user would like simulations which take into account as many physical effects as possible, most users would prefer not to need access to a computing cluster and/or support staff in order to run the simulations. For many use cases the interaction of photons with the various media in the instrument are of little importance and quick results take preference. However other user groups will indeed be interested in the difference in the signal to noise (S/N) ratio for different detector read-out schemes.

In order to provide a tool which meets the largest number of needs of the above mentioned user groups, we set ourselves the following requirements for SimCADO. 

SimCADO should:

\begin{enumerate}
    \item run in a language common among the astronomical community,
    \item be quick to install and run on an average personal computer,
    \item not require that the user be familiar with the inner workings of MICADO,
    \item give the user control over the simulation parameters including those relating to the optical train,
    \item maintain a modular design that can be easily adapted to the many upcoming design changes,
    \item use standard file types at the interface between the user and SimCADO.
    
\end{enumerate}



\begin{table}
    \centering
    \caption{Core and optional python dependencies for SimCADO}
    \begin{tabular}{| l|l | l|l |}
        \hline
        \multicolumn{2}{|c|}{Core dependencies} & \multicolumn{2}{c|}{Optional dependencies} \\
        Name  & Version  & Name  & Version  \\
        \hline
        \verb+numpy+ & \textgreater 1.10 & \verb+matplotlib+ & \textgreater 1.5 \\
        \verb+scipy+ & \textgreater 0.17 & \verb+poppy+ & \textgreater 0.4 \\
        \verb+astropy+ & \textgreater 1.1 & \verb+photutils+ & \textgreater 0.2 \\
        \hline
    \end{tabular}
    \label{tab:dependencies}
\end{table}

There are two main reasons behind the decision to restrict SimCADO to the architecture of a personal computer. Firstly, if the simulation effort is centrally managed and run, there is the possibility for a large bottle neck to occur in the work flow. 
If instead, the user can run SimCADO on her or his own laptop, she or he would be able to produce results in a much more timely manner. Furthermore, by decentralising the simulation effort, the amount of science cases that can be tested is no longer limited by the man-power or hardware available to the simulation team. 

\subsection{Imaging vs Spectroscopy}

MICADO is the first-light wide-field imaging camera for the E-ELT, and as such its main purpose is to provide imagary of the sky. While MICADO does offer a long-slit spectrographic mode, the priority for the MICADO design is to provide diffraction-limited imaging over the field of view. We have also adopted this approach. The current version of SimCADO only provides functionality for imaging in the wide-field and zoom modes. 

\section{Data Model behind SimCADO} 
\label{sec:data_model}

\begin{figure}
    \centering
    \includegraphics[width=0.9\textwidth]{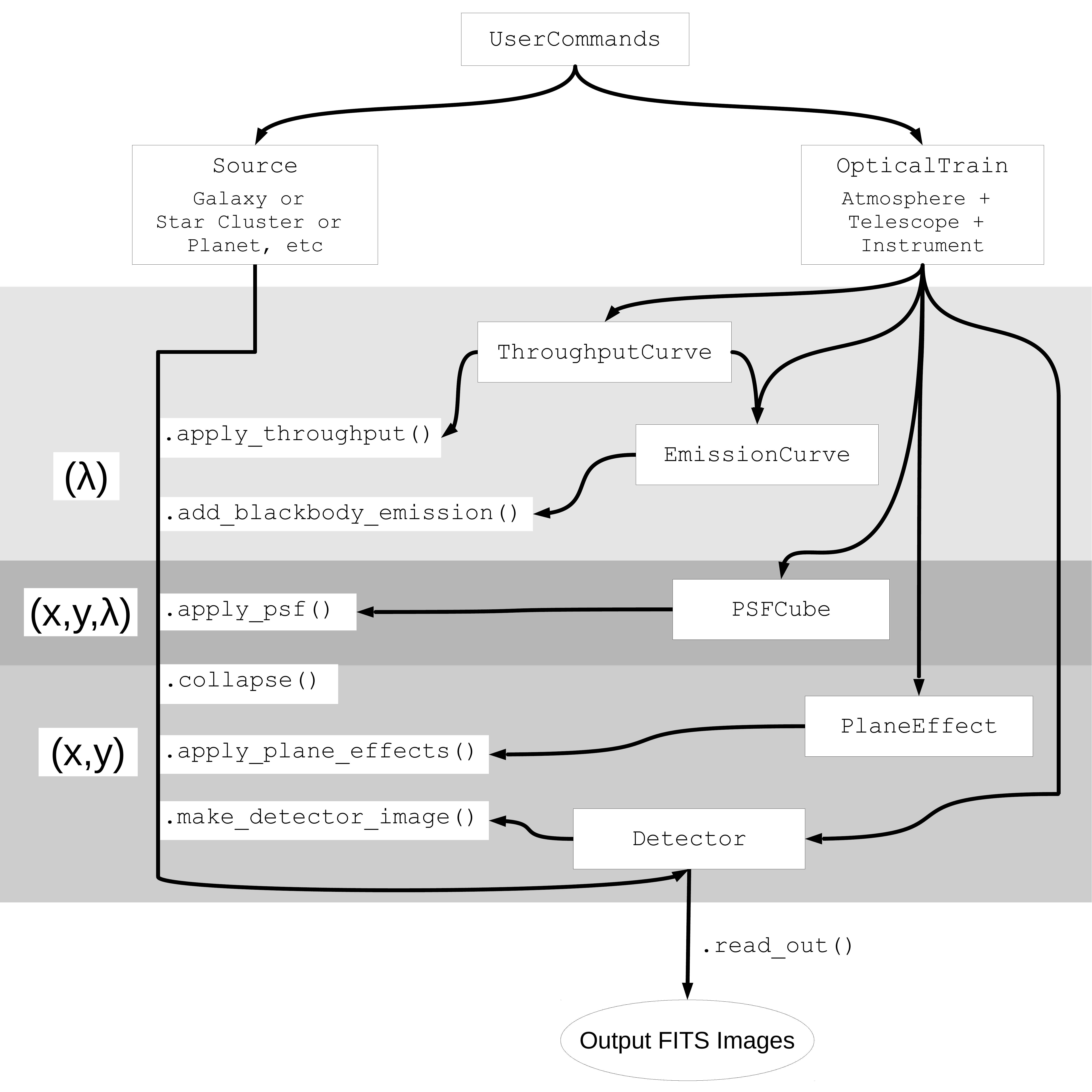}
    \caption{The simple version of the data model for SimCADO. Each of the boxed CamelCase words represents a class in SimCADO. The arrows show how each class interacts with the others. Words beginning with a dot are pseudo code functions to describe what is happening at that point during a SimCADO simulation run. An instance of the UserCommands class controls how each element in an instance of the OpticalTrain class is applied to a Source class object. The resulting image is held in a instance of the Detector class. How this process happens is described in section \ref{sec:opt_train_explain}}
    \label{fig:data_model}
\end{figure}

While designing SimCADO, we identified four main aspects involved in simulating MICADO detector array images. These were:
\begin{itemize}
    \item A representation of the source of photons
    \item A model of the optical path
    \item A description of the detector
    \item A list of parameters needed to control the simulation
\end{itemize}

The primary user interface (UI) of SimCADO involves class objects representing each one of these aspects. Running a simulation involves combining the data held within each of the four data structures in a manner representative of the whole optical system - See Figure \ref{fig:data_model}. This can either be executed automatically or manually, depending on whether the user wishes to view the intermediate data\footnote{Intermediate data in SimCADO does not refer to intermediate positions along the optical train, rather to how the distribution of photons on the focal plane changes at intermediate stages during a simulation after certain optical effects have been applied. E.g. after the transmission curves have been applied, but before the PSF has been taken into account.} or is simply interested in the final output.

\subsection{A note on SimCADO's internal representation of spectral data}

Before the main SimCADO classes are introduced, it is important to mention that the SimCADO data model does not use a data-cube (x,y,$\lambda$) to represent the photons travelling through the optical train. Instead the objects to be ``observed'' are converted into a list of point sources which are described by four values: two-dimensional coordinates on the plane of the sky, a reference to a unique spectrum and a weighting (i.e. a scaling of the intensity) for that instance of the spectrum. The decision to move away from the traditional data-cube model was motivated by both the need to reduce the amount of RAM required as well as reduce the amount of operations executed during a simulation. For example, a single MICADO detector array image contains nine 4k chips, each with $\sim$16  million pixels. If each pixel is $4\times$ oversampled in each direction and 4 byte integers are used, the size of the resulting image in memory will be:

\begin{equation}
    9 \times (4096 \times 4096) \times 4^2 \times 4 \text{ Bytes } \approx 10 \text{ GB}
\end{equation}

Note that this is a single oversampled monochrome detector array image. If the data-cube were to also include more than one wavelength, its size in memory would increase proportionally to the number of spectral channels. In the case of a K-band image of a field of stars with spectra taken from the Pickles (1998)\cite{pickles} library ($\delta\lambda$=0.5nm) the data-cube would need to contain over 800 layers. The resulting memory requirements would be $\sim$8 TB, far exceeding the capacity of the current generation of high-end laptops. By realising that for the most part, space is rather empty, and that therefore the majority of a data cube is also empty, SimCADO is able to circumvent this memory problem. The internal representation of photon sources by a pair of coordinates and a reference to a spectrum allows all the empty space, which would otherwise be represented in memory by zeros, to be ignored. Furthermore, in many science case simulations there is a large redundancy with regards to the number of unique spectra used. By only using references to the unique spectra, SimCADO further saves on memory and computational operations. The result of this vector-like representation is a vast reduction in computation time. Section \ref{sec:opt_train_explain} describes how SimCADO manipulates data in this form.


\subsection{The main classes of the SimCADO user interface}

The four class objects which represent the above-mentioned aspects are described in the following sections in greater detail. They are available in the upper level of the SimCADO package, i.e. \verb+simcado.OpticalTrain+ or can be called from their respective modules, i.e. \verb+simcado.optics.OpticalTrain+.

\subsubsection{Source}
	
The \verb+Source+ class is used to represent photon sources with a position and a spectrum. Examples include light from a star, galaxy or a planet. The class contains a list of unique spectra, a list of coordinates for each photon source, and a list of references which match each photon source to a spectrum in the list of spectra. The advantage of using this approach is that objects with highly similar spectra can both reference the same list position, thereby reducing the number of spectra that need to be manipulated during a simulation. 

An example of the benefit of this approach is the simulation of observations to map the star formation regions in galaxies. In SimCADO a galaxy is represented by an oversampled grid (typically $4\times$ the detector pixel scale) of ``point sources''. Much like in real observations, the spectrum for each grid element is the summation of the spectra for each source of photons (e.g. stars, nebula, etc) contained within the area of that grid element. For a basic example of a use case scenario, one can say that a typical galaxy contains a population of old stars (\textgreater1~Gyr) with an almost homogeneous spatial distribution, and a population of young stars (\textless1~Gyr) which exist in clusters scattered throughout the galaxy. Each grid element has its own unique position, yet elements that only contain an older stellar population all reference the spectrum for a ``generic'' older population. The same can be true for the younger population. The advantage in terms of computation time by exploiting such redundancy in the spectral domain is obvious.

In keeping in line with an object oriented methodology, the \verb+Source+ class contains a variety of methods to manipulate the internal data, and/or generate and return images of the source object. For a full list of the class methods, the reader is directed to the SimCADO documentation, however here is a brief list of the most useful methods in the \verb+Source+ class:

\begin{itemize}
    \item \verb+photons_in_range(+$\lambda_1,\lambda_2$\verb+)+ extracts the number of photons in a given wavelength range, $[\lambda_1,\lambda_2)$, from each spectrum.
    \item \verb+image_in_range(+ $\lambda_1,\lambda_2$\verb+, psf)+ generates a two dimensional image for a specific wavelength range, $[\lambda_1,\lambda_2)$, based on a supplied \verb+psf+ kernel.
    \item \verb+apply_optical_train(OpticalTrain)+ models the effects that the supplied \verb+OpticalTrain+ object will have on photons being emitted by the sources (e.g. PSF, transmission, jitter, etc). An image is generated for how the photons would be distributed on each chip in the focal plane array before they are converted to photo-electrons in the detector chips. These images are then passed onto the \verb+Chips+ contained in a \verb+Detector+ object (see below).
\end{itemize}

It should be noted that a \verb+Source+ object will not hold any images. Rather its internal methods generate images and pass these to the \verb+Detector+ objects. Internal images are generated on an oversampled detector pixel grid. The oversampling factor is an attribute of the optical train and can be set by the user.

\subsubsection{OpticalTrain}

As the name suggests, the \verb+OpticalTrain+ contains all the information needed to model the effects of the entire optical path on the incoming photons from sources held within a \verb+Source+ object. The optical train in this sense covers everything between where the photons where ``emitted'' and when they are converted into photo-electrons in the chips on the detector array. For the preset MICADO optical train, this includes: the atmosphere, the E-ELT, the MAORY\cite{maory} instrument (optional, depending on AO mode), all aspects of the MICADO optical configuration and the MICADO detector array. For the elements being developed outside the MICADO consortium, i.e. the E-ELT and MAORY, SimCADO currently uses the information made public during the respective phase A studies. For the elements being developed by the MICADO consortium, SimCADO combines information garnered from the latest consortium-internal documents. Table \ref{tab:elements} lists which elements of the atmosphere+E-ELT+MICADO optical train are currently simulated by SimCADO. 

\begin{table}[ht]
    \centering
    \caption{Elements of the optical train emulated by SimCADO. * These items currently use very basic models. The models are expected to improve as the detectors are tested. Section \ref{sec:modelling} describes how these elements are represented in SimCADO in detail.}
    \begin{tabular}{| l|l|p{0.6\textwidth} |}
        \hline
        Element  & Dimension & Effects \\
        \hline
        Atmosphere & $\lambda$ & Transmission, Emission \\
        E-ELT & (x, y, $\lambda$) & Mirror transmission and emission, vibration, tracking errors \\
        AO-corrected PSF & (x, y, $\lambda$) & MCAO, SCAO, No AO \\
        Entrance Window & $\lambda$ & Transmission \\
        Filter Set & $\lambda$ & Transmission \\
        Optical Design  & (x, y, $\lambda$) & Distortion, Transmission \\
        ADC & (x, y, $\lambda$) & Wavelength dependent shifts \\
        Derotator & (x, y) & Field rotation \\
        Detector & (x, y, $\lambda$) & Quantum efficiency, detector noise, persistence*, cross-talk* \\
        \hline
    \end{tabular}
    \label{tab:elements}
\end{table}

When initialised, an \verb+OpticalTrain+ object reads in all the information necessary to model the optical path. This can either be read in from a pre-prepared optical train file, or from files which relate to individual aspects of the optical train (e.g. transmission curves in ASCII format, or FITS images with PSF kernels, etc.). 

As the effect of the transmission curves of each element is cumulative, \verb+OpticalTrain+ combines all transmission curves internally into a ``master'' transmission curve
\footnote{OpticalTrain actually creates three ``master'' transmission curves - one for each of the major sources of photons: 

1) tc\_source - equivalent to the throughput for the astronomical source. It contains all transmission curves, i.e. the atmosphere, E-ELT mirrors, and transmission through MICADO

2) tc\_atmo - equivalent to the throughput for the atmospheric photons. It contains all transmission curves minus the atmospheric transmission , i.e. E-ELT mirrors and transmission through MICADO

3) tc\_mirror - equivalent to the throughput for grey-body photons from the primary mirror - It contains only the total throughput of the MICADO instrument}. 
It is this ``master'' transmission curve which is then applied to the spectra in a \verb+Source+ object. 

\verb+OpticalTrain+ currently assumes that the two major sources of background photons - the atmosphere and the primary mirror - can be taken as approximately spatially constant\footnote{For the case of a single detector readout, the assumption of a constant sky background is valid, however, over the course of a night variations in the NIR background have been found - Pedani (2014)\cite{pedani14} and Moreels (2008)\cite{moreels08}. In fuure releases of SimCADO we will aim to address this issue.} with an associated spectrum. In the case of the atmosphere, this emission spectrum is taken from \verb+SkyCalc+\cite{noll2012, jones2013}. Because of the assumed spatial constancy, only transmission effects apply to the background photons. As such, \verb+OpticalTrain+ immediately calculates and stores the average number of background photons through the specified filter bandpass.

The PSF kernels used for the two major AO modes (MCAO and SCAO) are kept in FITS files delivered with the package so that they can be easily updated. They currently do not change over the field of view. Field-varying PSF kernels will be implemented once updated kernels are available from the MCAO and SCAO teams. The user may also use their own PSF kernel in the form of a FITS image file, providing the correct FITS keywords are in the header, or generate analytical PSFs with SimCADO's built-in functions (See Section \ref{sec:psf}).

Other wavelength-independent effects, such as field rotation or telescope jitter, are applied directly through built-in functions. The required parameters are parsed through the \verb+UserCommands+ object required to initialise an \verb+OpticalTrain+ object.

\subsubsection{Detector}
	
Although only a single element in the whole optical train, the detector is the final interface between the incoming photons and the observer. The detector chips are the primary source of systematic effects in raw images, therefore it is very important to both understand, and be able to model the detector chip characteristics accurately. 

The \verb+Detector+ class contains a list of \verb+Chip+ objects and the layout of the chips in the focal plane. It also holds information about how the chips are to be read. Each \verb+Chip+ object contains the information about its own noise properties. When the focal plane image of a \verb+Source+ is created, the image is re-binned and passed straight onto the relavent chips. For quick simulations, the \verb+.readout()+ method samples the signal on the chips only once (at the end of an exposure) and applies a Poisson noise distribution to the expected signal. If simulation time is not an issue, a full up-the-ramp read-out sequence is used. As the time between non-destructive read-outs (typically $\sim$3 sec) is generally small compared with the length of a single exposure, the user should not be surprised by extended simulation times.

Similar to the \verb+OpticalTrain+, a \verb+Detector+ object can be created from a series of individual FITS image files representing the noise properties of the chips, and an ASCII table describing the layout of the chips on the detector plane, or it can be read from a pre-prepared detector file. 

\subsubsection{UserCommands}
By default SimCADO uses parameter values that correspond to the MICADO instrument. For ease of use (i.e. so that the user need not pass 20 parameters to every function), all parameters needed to run a simulation are defined and held in a python dictionary in the \verb+UserCommands+ class. Also included in the \verb+UserCommands+ class are frequently used quantities and vectors, e.g. vectors with the wavelengths of the edges and centres of each wavelength bin used when integrating over the number of photons in a filter bandpass. 


Following tools like SExtractor\cite{sextractor}, a \verb+UserCommands+ object can also be initialised from an ASCII file containing keyword-value pairs. The user only needs to list the keyword-value pairs that they want to change - the rest are set to the default values. A \verb+UserCommands+ object is used to set up instances of the two main hardware classes: \verb+OpticalTrain+ and \verb+Detector+. Attribute values can be updated on-the-fly, although any \verb+OpticalTrain+ or \verb+Detector+ objects that were created before the change will need to be re-initialised.

\subsection{An short example of using the main SimCADO classes}

A simple simulation with SimCADO requires very little coding. For example, if we would like to simulate the detector output for a single exposure of a $10^4M_{\odot}$ open cluster in the Large Magellanic Cloud (d$\sim$50 kpc), we first create a \verb+UserCommands+ object with all the default values, and use it to generate an \verb+OpticalTrain+ object and a \verb+Detector+ object.

\begin{verbatim}
    >>> import simcado
    >>> cmds = simcado.UserCommands()
    >>> opt_train = simcado.OpticalTrain(cmds)
    >>> fpa = simcado.Detector(cmds)
\end{verbatim}

Next we create a \verb+Source+ object for the open cluster using a convenience function from the \verb+optics_utils+ module. We call the method \verb+.apply_optical_train()+ and pass both the \verb+OpticalTrain+ and \verb+Detector+ objects. This method is the heart of the simulation, as it converts the lists of point sources held in \verb+Source+ into a two dimensional distribution of expected photon flux. Depending on the parameters, i.e. the FoV or spectral resolution required, this step can take anywhere from 10 seconds to 10 minutes. During this process, the ``imagery'' generated inside the \verb+Source+ object is transferred to the \verb+Detector+ object. 

\begin{verbatim}
>>> src = sim.optics_utils.source_1E4_Msun_cluster()
>>> src.apply_optical_train(opt_train, fpa)
\end{verbatim}

The raw detector array images can then be read out by calling the \verb+Detector+'s \verb+read_out+ method. If no filename is specified, the method returns an astropy \verb+HDUList+ with a FITS image extension for each chip in the detector array.  

\begin{verbatim}
>>> fpa.read_out(filename="my_raw_image.fits")
\end{verbatim}

\begin{figure}
    \centering
    \includegraphics[width=0.98\textwidth]{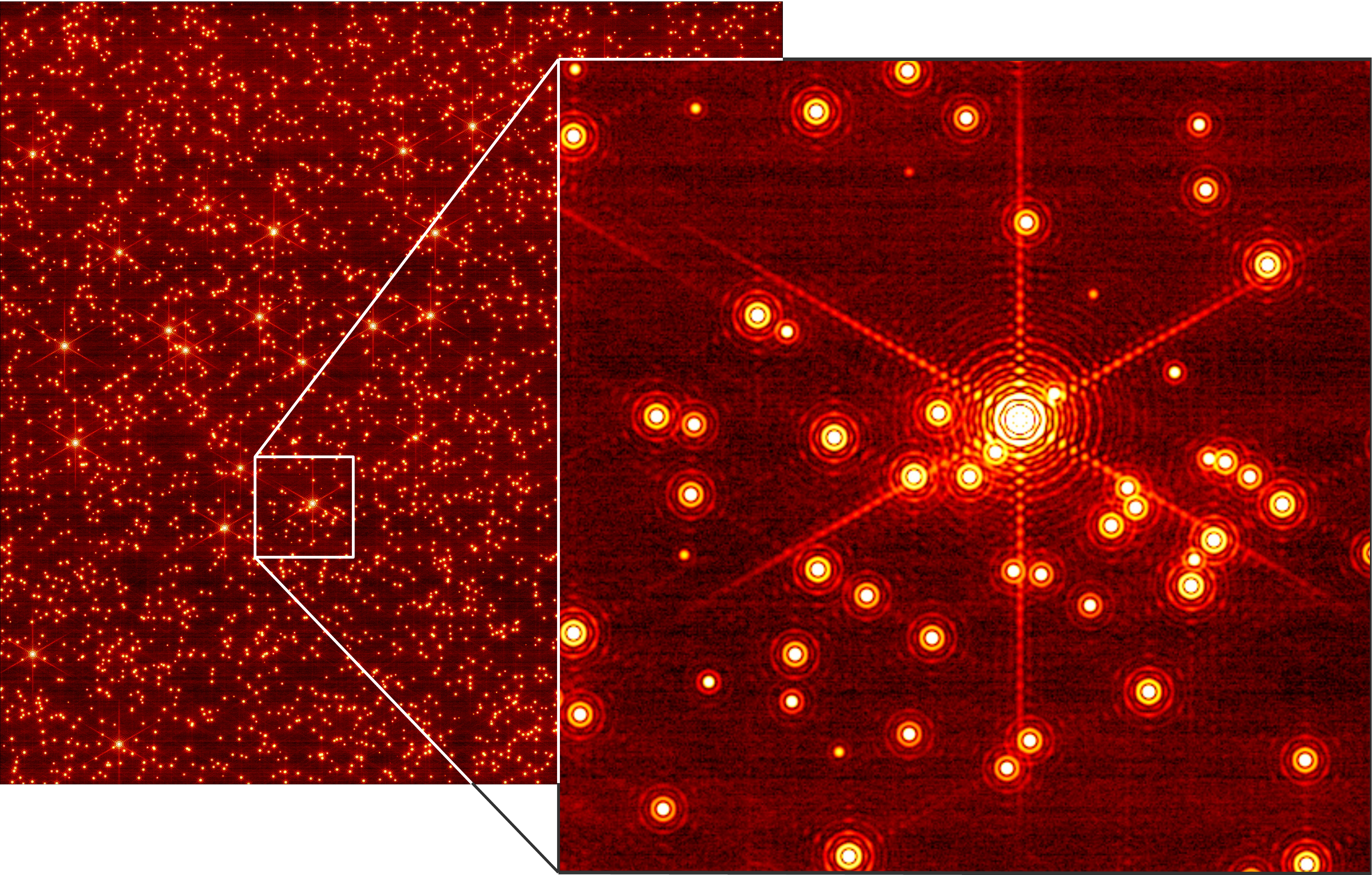}
    \caption{The heart of a compact star cluster in the LMC as seen with SimCADO. The left image shows the central H4RG chip in the 3$\times$3 detector array, corresponding to a FoV of $\sim$16 arcsec, or a linear distance of 0.25 pc. The right image shows a 512 $\times$ 512 pixel section of the image. For this simulation the atmospheric effects were removed so that both the diffraction pattern of the E-ELT's primary mirror as well as the detector noise from the H4RG chips (courtesy of Bernhard Rauscher's HxRG Noise Generator\cite{nghxrg}) could be seen.}
    \label{fig:star_cluster}
\end{figure}

Figure \ref{fig:star_cluster} shows the results of a simple simulation similar to that described in this section. 

\subsection{Method behind applying an OpticalTrain}
\label{sec:opt_train_explain}
\begin{figure}
    \centering
    \includegraphics[width=0.98\textwidth]{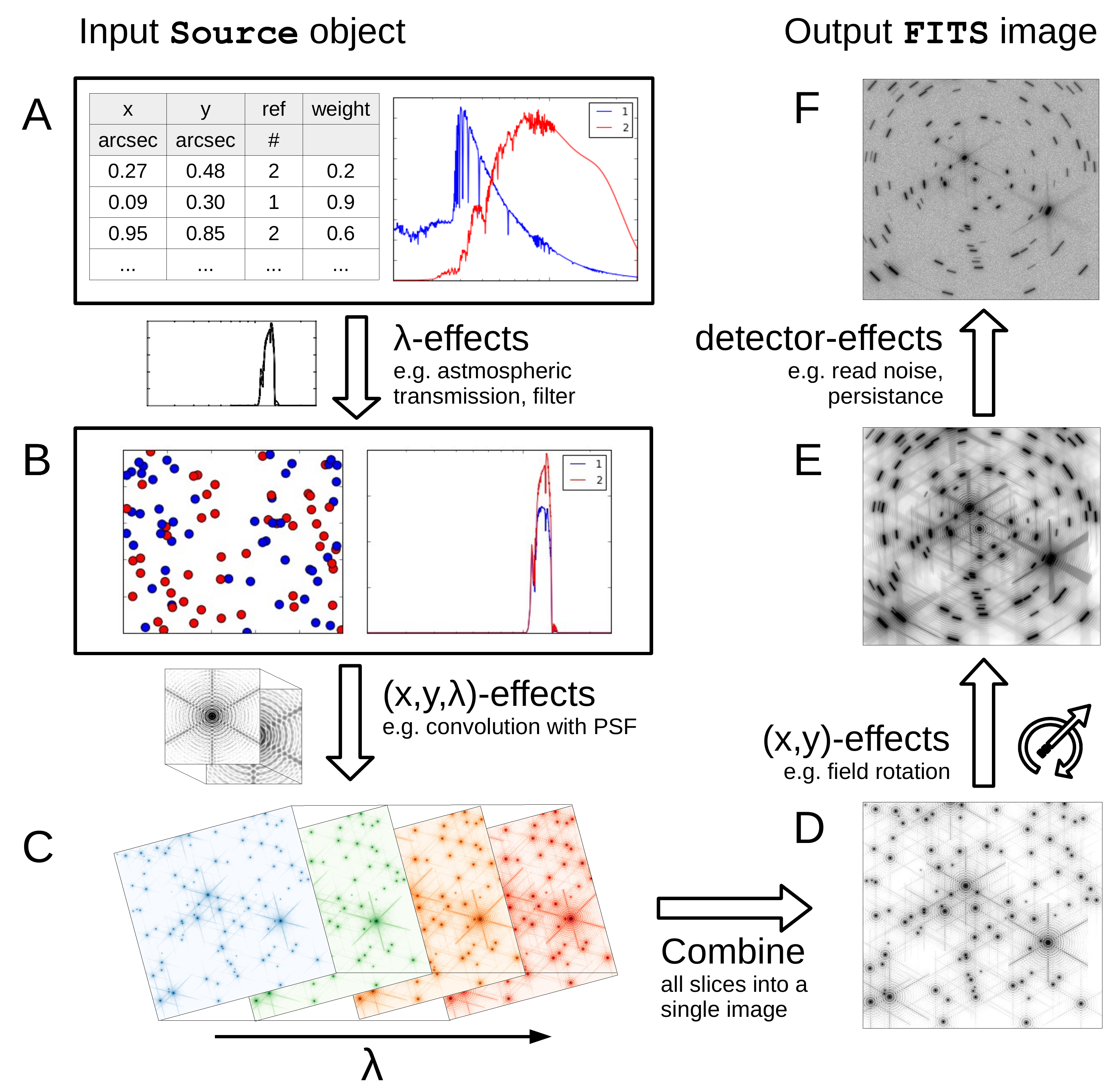}
    \caption{A simplified graphical representation of what happens to a Source object when the method \mbox{\textless Source\textgreater .apply\_optical\_train()} in called. The diagramme is explained in detail in Section \ref{sec:opt_train_explain}. The flow of events can be summarized so: (A) a Source object for a mock open cluster is created.  All effects of the optical train which only act in the wavelength-domain ($\lambda$-effects) are applied to the spectra in the array. (B) Several images of the spatial extent of the Source are created and weighted by the number of photons in each low resolution spectral bin. Any elements that affect the incoming photons in all three dimensions (x,y,$\lambda$) are taken into account here (C) before (D) the images are added together. Purely spatial effects (e.g. telescope vibrations) are then applied to the (E) monochrome image and detector noise is added. (F) The resulting image represents the spatial distribution of all photo-electrons (from source object + atmosphere + primary mirror) plus the electronic noise generated by reading out the detector chips. A FITS file is either written to disk or returned to the used with an image extension for each chip in the detector array.}
    \label{fig:apply_optical_train}
\end{figure}

In an ideal world, SimCADO would apply all spectral and spatial changes at the resolution of the input data. However, as previously mentioned, the memory requirements to do this are well outside the limits of a personal computer. The solution to this problem is to split the effects based on dimensionality. For certain elements in the optical train, the spectral and spatial effects can be decoupled (e.g. purely transmissive elements like the filters versus purely spatial effects like telescope vibration). For other elements, most notably the PSF and ADC, all three dimensions must be considered simultaneously. When applying an \verb+OpticalTrain+, SimCADO follows the procedure described graphically in Figure \ref{fig:apply_optical_train}. The example used in Figure \ref{fig:apply_optical_train} is for a simplified stellar cluster with only two different stellar types - A0V and K5V type stars. 

\begin{enumerate}[label=(\Alph*)]
    \item The original \verb+Source+ object includes an array of spectra for each {\bf unique} photon source (in the case of Figure \ref{fig:apply_optical_train}, there are only two unique spectra) and four vectors: \verb+x+, \verb+y+, \verb+ref+, \verb+weight+. \verb+x+ and \verb+y+ hold the spatial information for each photon source, \verb+ref+ connects each source to a spectrum in the array of spectra and \verb+weight+ allows the spectrum to be scaled. 
    
    \begin{itemize}
        \item $\lambda$-effects. The first step in \verb+apply_optical_train()+ is to combine all optical elements which only act in the wavelength domain (e.g. filters, mirrors, etc.) into a single effect, then apply that effect to the array of spectra in the \verb+Source+ object. 
    \end{itemize}

    \item The spectra in the \verb+Source+ object are now representative of the photo-electron count at the detector, assuming a perfect optical train and at the internal spatial resolution of the simulation, i.e. \textit{not at the pixel scale of the detector}. The position vectors are converted into a two-dimensional ``image'' of the \verb+Source+.
    
    \begin{itemize}
        \item (x,y,$\lambda$)-effects. The second step includes creating ``slices'' through the data. The spectra are binned according to several criteria (ADC shift, PSF FWHM difference, etc) with a spectral resolution anywhere from R=1 to R\textgreater100, and the number of photons per source in each wavelength bin is calculated. The sources in each ``slice'' are scaled according to the number of photons in each bin. The relevant spatial effects (atmospheric dispersion, convolution with PSF kernel, etc.) are then applied to each slice in turn.
    \end{itemize}        
    
    \item At this stage, the \verb+Source+ object contains many spectral slices. Each is essentially the equivalent of a (\textit{very}) narrow-band filter image.
    
    \item All spectral effects have been taken into account, and so the binning in the spectral domain is no longer needed. The third step in \verb+apply_optical_train()+ is to add all the slices together to create a single monochrome image.
    
    \begin{itemize}
        \item (x,y)-effects. Fourth in the series of operations is to apply the purely spatial effects (e.g. telescope jitter, field rotation, etc) to the monochrome image. 
    \end{itemize} 
    
    \item The resulting image represents how the incoming photons from the source would be distributed on the focal plane after travelling through the entire optical train. At this point the background photons are also added to the image. Because SimCADO doesn't take into account the changing sky background, the sky emission is approximated as a constant background photon count determined from an atmospheric emission curve (either provided by the user or generated by \verb+SkyCalc+\cite{noll2012,jones2013}). The mirror blackbody emission is also approximated as spatially constant. For all filters, with the exception of K, the amount of additional photons due to the mirror is close to negligible.
    
    \begin{itemize}
        \item detector-effects. The image is resampled down from the internally oversampled grid down to the pixel scale of the detector chips - in the case of MICADO either 4 mas or 1.5 mas, depending on mode. The final step is to add noise in all its forms to the image. Various aspects of the detector noise (correlated and uncorrelated white and pink noise read-out (see Rauscher 2015\cite{nghxrg}), dead pixels, etc.), as well as photon shot noise for both the atmospheric and object photons are taken into account. Further effects (e.g. detector persistence, cross-talk, etc) are also added to the image at this point.
    \end{itemize}   
    
    \item The final image represents the spatial distribution of all photo-electrons (from the source object + atmosphere + primary mirror) plus the electronic noise generated by reading out the detector chips. The images from all the chips considered in a simulation are packed into a FITS extension and the FITS file is either written out to disk, or returned to the user if generated during an interactive Python session.

\end{enumerate}

\subsection{Generating input for simulations}

As the data format used by the \verb+Source+ class is not standard, SimCADO contains a module to convert more standard data types into \verb+Source+ objects. Currently the SimCADO module \verb+source_utils+ offers the following ways to create a \verb+Source+ object:

\begin{itemize}
    \item from a FITS cube,
    \item from a FITS image plus an ASCII spectrum,
    \item from an ASCII list of coordinates with references and an ASCII table of spectra,
    \item from a combination of several \verb+Source+ objects,
    \item by using the in-built functions in \verb+simcado.source_utils+ to generate \verb+Source+ objects
    \item by passing a series of arrays already in an iPython Notebook when initialising a \verb+Source+ object
\end{itemize}
    
\verb+Source+ objects can also be saved to disk for later use.

The other input required for SimCADO is the data on the optical train. This can either be supplied in the form of ASCII and/or FITS files which detail all the aspects the user wishes to include in the simulation, or in the form of pre-prepared \verb+OpticalTrain+ and \verb+Detector+ files. 

\subsection{Modules behind the scene}
\label{sec:psf}

In order for SimCADO to function, several other modules and classes need to be mentioned.	
\begin{description}
    \item \verb+simcado.psf+
    
    This module contains the classes \verb+PSF+ and \verb+PSFCube+. \verb+PSFCube+ holds a list of \verb+PSF+ objects and provides the functionality to manipulate the PSF kernels. The most commonly used class is \verb+UserPSFCube+, which reads in a FITS file containing one or more PSF kernels. The module also provides functions to generate a variety of ideal case analytical PSFs, including Gaussian, Airy and Moffat PSF kernels. 
    
    \item \verb+simcado.spectral+
    
    \verb+spectral+ contains the classes used to represent transmission and emission curves. The base class is \verb+TransmissionCurve+ and holds, among others, two arrays: \verb+.lam+ and \verb+.val+ which hold the information about the wavelength and corresponding degree of transmission. By providing units and removing the normalisation, \verb+TransmissionCurve+ can be extended to accommodate an \verb+EmissionCurve+ with units of [ph/s/bin]. 
    
    \item \verb+simcado.spatial+
    
    The \verb+spatial+ module contains no class data structures, but rather a series of functions that mimic the purely spatial effects of elements in the optical train. Examples include field rotation, tracking errors and shifts. These effects can either be implemented in a purely linear fashion, for example simulating if the telescope tracking was turned off completely, or in a Gaussian manner, which is instead used to simulate uncertainty in the centre of the tracked position.
    
\end{description}

\section{Modelling the Optical Path}
\label{sec:modelling}

As mentioned in the Section \ref{sec:opt_train_explain}, SimCADO makes use of the fact that not all elements affect the incoming photons in all dimensions. Filters, for example, strongly alter the incoming spectrum of light, but have a negligible effect on the spatial distribution of the photons. A faulty Derotator on the other hand, has a noticeable effect on how sharp point sources are, however the effect is equal across all wavelengths. As such the effects of each of these two elements can be applied independently of the other to the incoming photons. In the following section, these effects are referred to as spectral (also $\lambda$-, or 1D-) and spatial (also (x,y)-, or 2D-) effects. The latter include, but are not limited to; shifts, rotations, convolutions and distortions. There are also elements which affect both the spatial and spectral nature of the incoming photons, e.g. the point spread function (PSF) of the telescope. These elements are referred to as spectrospatial (also (x,y,$\lambda$)-, or 3D-) elements. 

\subsection{Representing elements in the optical train in SimCADO}

The following section describes how each element in the optical train is represented in SimCADO:

\begin{itemize}
    \item Atmosphere - ($\lambda$) 
    
    The atmosphere is the element in the optical train that has the largest effect on the incoming photons, in all three dimensions. However, because of the E-ELT's adaptive optics facilities, the spatial effect of the atmosphere (namely the seeing-limited PSF) will mostly be removed, with the residual being combined with the PSF of the E-ELT. Both the MAORY MCAO\cite{maorySim} and MICADO SCAO\cite{clenet2013} teams are in the process of simulating post-AO PSFs, therefore for the purpose of SimCADO only the spectral aspects of the atmosphere need be addressed. Detailed simulations of the transmission and emission curves for the atmosphere have been generated by the \verb+SkyCalc+ tool\cite{noll2012,jones2013} and are used as the default in SimCADO.
    
    \item Telescope - (x,y,$\lambda$)
    
    There are many aspects on the E-ELT that need to be addressed by SimCADO, each of which is active in a different subset of the three dimensions. These include, but are not limited to: 
    \begin{enumerate}
        \item the reflectance of the mirrors - ($\lambda$),
        
        SimCADO assumes that the same coating will be used for all 5 mirrors. Currently SimCADO uses the reflectance curve provided by ESO for a layered silver-aluminium (AgAl) coating with a magnesium fluoride (MgF$_2$) protective layer\cite{boccas06}\footnote{\url{https://www.eso.org/sci/facilities/eelt/science/drm/tech_data/telescope/}}.
        
        \item the thermal emission of the mirrors - ($\lambda$),
        
        A simple grey-body curve is used to estimate the background photon flux due to the 5 mirrors in the system.
        
        \item the AO corrected PSF - (x,y,$\lambda$), 
        
        Currently SimCADO provides the option to use several obscured types of PSF kernels: analytical PSFs generated on-the-fly which follow the form of a Gaussian, Airy or Moffat function (see Section \ref{sec:psf}; simulated diffraction limited PSFs for a 39m mirror with hexagonal segments using the POPPY package\cite{poppy}; or external PSFs from the SCAO or MCAO simulation efforts. If the PSF kernels are generated internally, the spectral resolution of the PSFs is adjusted to that of the simulation. If an external FITS file containing PSF kernels is provided, the PSF nearest to the needed wavelength is chosen (although the user is warned if the difference in wavelength is \textgreater10 nm.)
        
        \item the residual vibration of the telescope structure - (x,y),
        
        Although the AO systems will be able to adapt to the low frequency vibrations, vibrations with frequencies higher than the wavefront sensor (WFS) read-out time will continue to affect the image quality. As vibrations are an artifact of the telescope structure, and not the optics, they are wavelength independent. The vibrations also act on timescales much shorter than a single detector readout, therefore the sum effect of the telescope vibrations can be approximated as an additional convolution with a Gaussian PSF. Currently, SimCADO uses a circular Gaussian kernel, however we will update this model once more information on the vibrational modes of the main E-ELT structure becomes available. 
        
        \item the tracking uncertainties - (x,y) 
        
        With a pixel-scale of 4mas, tracking an object across the sky becomes rather challenging. During a single detector read-out ($\sim$3 sec) the sky will have moved almost a full MICADO field of view. This means the tracking system will have updated the position of the image more than 25,000 times during this time. It is improbable that the image position will stay exactly in the centre of the frame during each movement\footnote{The AO WFS units will only have read-out $\sim$750 frames in in the same time period. Although the AO systems may be able to correct for tracking errors, there will still be a residual shift due to these different time scales.}. Therefore SimCADO also has the functionality to blur the image along a specific vector, either using a Gaussian or linear distribution, in order to simulate errors in the tracking process. Again, once more information becomes available regarding the tracking systems, this aspect will be updated accordingly.
        
    \end{enumerate}
    
    \item Instrument - (x,y,$\lambda$)
    
    The various elements of the MICADO instrument are represented as a collection of different effects on the incoming photons. As MICADO is still in the preliminary design phase, many of the elemental descriptions are expected to be updated as the instrument design is refined over the next 4 years. The elements modelled by SimCADO are:
    \begin{enumerate}
        \item the cryostat window - ($\lambda$)
        
        Currently set to 95\% to reflect the current choice of material for the cryostat window. The material exhibits a flat transmission curve through out the NIR wavelength range.
        
        \item the collimator and camera optics (x,y), ($\lambda$)
        
        The current design of MICADO contains nine reflective surfaces which affect both the spectral and spatial characteristics of the incoming photons. However, the spatial effect can be decoupled from the spectral effect due to the purely reflective nature of the optics. Thus the collimator+camera optics can be split into a spectral component based on the reflectance of the mirror surfaces and a spatial component based on the distortion map of the system.
        
        \item Atmospheric Dispersion Corrector (ADC) - (x,y,$\lambda$)
        
        The job of the ADC is to remove the atmospheric dispersion which leads to an elongation of point sources within a filter bandpass. If the ADC is performing as per specifications, there should be no noticeable elongation in the detector images, and hence this element need not be considered in the simulations. However, if the ADC performance is sub-par, then SimCADO adds in a shift along the zenith direction relative to the degree to which the ADC is misbehaving. If the ADC is taken out of the optical train completely, the shift that SimCADO adds to each spectral layer is equal to the full extent of the atmospheric dispersion for the wavelength of the layer at the zenith distance of the observation.
        
        \item Derotator - (x,y)
        
        In order to counteract the rotation of the sky, the entire MICADO cryostat is rotated as an observation takes place. As with the tracking of the E-ELT main structure, there will always be a certain amount of error induced because of the mechanical nature of the derotator. For photon sources near the centre of the FoV, these errors are negligible, however at the edge of the field, the shift become non-trivial. Similar to the ADC in SimCADO, if the derotator is performing perfectly, this effect need not be considered. However if it is turned off, SimCADO adds a rotational blur (see E in Figure \ref{fig:apply_optical_train}) to the focal plane image. This can be done linearly, i.e. no derotation, or with a Gaussian distribution, to simulate errors induced by uncertainty in the rate of rotation.

    \end{enumerate}    
    
    \item Detector - (x,y)
    
    The MICADO detector array will consist of nine Teledyne Hawaii-4RG detector chips\cite{hall2011}, similar to the ones used in the JWST NIRSpec Instrument\cite{rauscher07}. Although the chips are sensitive to a large range of wavelengths, they do not differentiate between photons of different energies. Hence they have a purely spatial effect on the output image. The majority of the work in modelling the noise characteristics of the Hawaii chips has already been done by Rauscher et al. (2012)\cite{rauscher12} and Moseley et al. (2010)\cite{moseley10}, which led to the development of the HxRG Noise Generator code\cite{nghxrg}. SimCADO uses this code to apply detector noise frames to the final chip read-out images.

\end{itemize}

\subsection{Aspects of the optical train to be included in later releases}
As time is of the essence during the design phase, we have decided to release the SimCADO core package to certain members of the MICADO consortium. This serves the dual purpose of allowing these members to become familiar with SimCADO, as well as helping us to improve the usability of the code. The core package allows basic simulations of the full MICADO detector array, which will be suitable for the majority of use cases. In future releases for the imaging module of SimCADO we aim to include the following features:

\begin{itemize}
    \item Observational coordinates - The ability to specify the R.A. and Dec. coordinates, as well as the date and time of any observation will allow various science cases to be tested for specific objects, e.g. solar system bodies, star clusters in nearby galaxies (e.g. NGC300).
	
	\item Extra-terrestrial optical path elements -  Zodiacal light, galactic extinction, atmospheric extinction and scattered moonlight (bluewards of 1 \textmu m) also contribute a non-negligible amount to the background photon flux and should therefore also be included in science case studies
	
	\item Variable sky background -  As the sky background in the NIR is dominated by OH lines, and the OH density varies over time along the line of sight, the varying sky background emission can also play a large role in determining the sensitivity of long exposure observations.
	
	\item Missing segments - According to the plan B construction scenario for the E-ELT, several segments will be missing from the primary mirror each night due to the need to re-coat the mirror surface. This will no doubt affect the PSF.
	
	\item PSF variability over the FoV - Currently SimCADO convolves the source image with a single wavelength-dependent PSF. However the PSF at any point in the FoV will depend on the distance to the guide stars being used for the AO correction.
	
	\item Updated instrumental distortion map - As the optical design of MICADO progresses, so too will the estimation of the instrumental distortion. For the majority of science cases this effect is negligible, however for astrometric science cases, as well as the data reduction process, the distortion of the pixels over the FoV is very important. 
	
\end{itemize}
\section{Conclusion} 
\label{sec:outlook}

SimCADO currently contains all the functionality needed to generate realistic mock detector array readout images for the MICADO instrument on the E-ELT. By being written in Python 3 and only requiring \verb+numpy+, \verb+scipy+ and \verb+astropy+ as the core dependencies, we have met the first and second of our development goals. The third and fourth goals are met by SimCADO's API. The user does not need to have any prior knowledge of the MICADO instrument to use SimCADO, yet through the \verb+UserCommands+ class the user still has full access to all the parameters used to control simulations, if she or he so desires. The \verb+Source+, \verb+OpticalTrain+ and \verb+Detector+ classes allow for simulations to be run with as little as five lines of code, yet also give access to all the internal mechanisms governing the flow of the simulation. Finally SimCADO fulfils our fifth development goal by accepting input in either the commonly used FITS or ASCII formats, and outputting all data as FITS files. By staying within the standard FITS framework, images generated by SimCADO can therefore be used by the veritable zoo of other programs already in use within the astronomical community.

Just as the primary function of MICADO is wide-field imaging, the current state of the SimCADO package allows the user to simulate mock imagery for the MICADO detector array. In a future release the ability to simulate the spectroscopic capabilities of MICADO will become available. The modularity of the SimCADO design will allow us to add this functionality with minimal effort. However simulations with SimCADO will only ever reflect the current status of the MICADO design. As MICADO matures, so too will SimCADO. Similarly, we will endeavour to incorporate more detailed information on the MAORY and E-ELT designs as it becomes available.

\appendix    

\acknowledgments     
This research made use of Astropy, a community-developed core Python package for Astronomy \cite{astropy}.
This research made use of POPPY, an open-source optical propagation Python package originally developed for the James Webb Space Telescope project \cite{poppy}. 
SimCADO incorporates Bernhard Rauscher's HxRG Noise Generator package for python\cite{nghxrg}. 
SimCADO makes use of atmospheric transmission and emission curves generated by ESO's SkyCalc service, which was developed at the University of Innsbruck as part of an Austrian in-kind contribution to ESO. 
This research is partially funded by the project IS538003 of the Hochschulraumstrukturmittel (HRSM) provided by the Austrian Government and administered by the University of Vienna.
The authors would also like to thank all the members of the consortium for their effort in the MICADO project, and their contributions to the development of this tool. 


\newcommand{\aap}{Astronomy \& Astrophysics}
\newcommand{\aaps}{Astronomy \& Astrophysics, Supplement}
\newcommand{\mnras}{Monthly Notices of the Royal Astronomical Society}
\newcommand{\procspie}{Proceedings of the International Society for Optical Engineering}
\newcommand{\pasp}{Publications of the Astronomical Society of the Pacific}
\newcommand{\apjs}{The Astrophysical Journal, Supplement}

\bibliography{report}   
\bibliographystyle{spiebib}   

\end{document}